\documentclass[11pt]{article}
\usepackage{hyperref}
\usepackage{textcomp}
\pdfoutput=1
\begin{document}
\title{High-Speed Acquisition of Free Vortex Formation}
\author{Ahmad Falahatpisheh, Arash Kheradvar \\
\\\vspace{6pt} Mechanical and Aerospace Engineering Department, \\ University of California, Irvine, Irvine, CA 92697, USA}
\maketitle
\begin{abstract}
The formation of a free-vortex has been captured by using a high-speed camera (Y3, IDTVision, Inc.). The experiment is conducted using a rectangular tank, which is filled with tap water. The water free surface is open to atmospheric pressure and is at room temperature, 25\textcelsius. Water occupies a volume of $25\times 25\times 10$cm$^3$. By using a stirring-spoon, the stagnant water is forced to rotate at a rate of $2\pi$/sec. Once all the points in the water is rotating, it will be drained from a ball valve, with a diameter of 5mm, from the bottom of the tank and the acquisition starts. The formation of the vortex is captured with a resolution of $352\times 824$ pixels at 200 frames per seconds (fps) and is exported at 5fps and with a resolution of $1280\times 720$ in a ``fluid dynamics video". The duration of the video in real time is 3.9 seconds. The slow motion video is 160 seconds. The height of the water remains almost unchanged while acquiring the images. 
\end{abstract}

\end{document}